\documentclass{article}

\usepackage{PRIMEarxiv}

\usepackage[utf8]{inputenc} % allow utf-8 input
\usepackage[T1]{fontenc}    % use 8-bit T1 fonts
\usepackage{hyperref}       % hyperlinks
\usepackage{url}            % simple URL typesetting
\usepackage{booktabs}       % professional-quality tables
\usepackage{amsfonts}       % blackboard math symbols
\usepackage{nicefrac}       % compact symbols for 1/2, etc.
\usepackage{microtype}      % microtypography
\usepackage{lipsum}
\usepackage{fancyhdr}       % header
\usepackage{graphicx}       % graphics
\usepackage{subcaption}
\graphicspath{{media/}}     % organize your images and other figures under media/ folder

\title{Reframing SAR Target Recognition as Visual Reasoning: A Chain-of-Thought Dataset with Multimodal LLMs
 \footnotemark
}

\author{
  Chaoran Li \\
  Shanghai Jiao Tong University\\
  \texttt{lcr0203@sjtu.edu.cn} \\
   \And
  Xingguo Xu \\
  Dalian University of Technology \\
  \texttt{xuxingguo@mail.dlut.edu.cn} \\
     \And
  Siyuan Mu \\
  Shanghai Jiao Tong University \\
  \texttt{hbmsy9@sjtu.edu.cn} \\
}

\begin{document}
\maketitle

\footnotetext{This work is in progress.}

\begin{abstract}
In the context of Synthetic Aperture Radar (SAR) image recognition, traditional methods often struggle with the intrinsic limitations of SAR data, such as weak texture, high noise, and ambiguous object boundaries. This work explores a novel perspective by reformulating SAR target recognition as a multimodal reasoning task. 
We leverage multimodal large language models (MLLMs), specifically GPT-4o, to perform target classification based on SAR imagery, guided by candidate categories and enhanced with Chain-of-Thought (CoT) reasoning. 
A new dataset is constructed based on the FAIR-CSAR benchmark, comprising raw SAR images, structured target annotations, candidate label sets, and GPT-generated CoT reasoning chains. 
Experimental results show that the MLLMs are capable of generating logically coherent and interpretable inferences in most scenarios. Our analysis highlights both the strengths and current limitations of MLLMs in interpreting SAR imagery, and we provide detailed insights into model behavior through failure case analysis. This work demonstrates the feasibility of incorporating MLLMs into SAR analysis pipelines and establishes a foundation for future research in SAR-oriented visual reasoning.

\end{abstract}

% keywords can be removed

\section{Introduction}
Synthetic Aperture Radar (SAR) imagery plays a significant role in various applications, including military reconnaissance, maritime surveillance, and disaster response. Compared to optical imaging, SAR offers all-weather, all-day imaging capabilities, making it effective in complex weather conditions and nighttime environments\cite{moreira2013tutorial}. However, due to the fundamental differences in imaging mechanisms between SAR and optical sensors, SAR target recognition presents numerous challenges, such as high noise levels, weak texture representation, and blurred object boundaries\cite{li2021comprehensive}. 
These characteristics make SAR target recognition a valuable but inherently difficult research problem.

In recent years, mainstream SAR image recognition methods have increasingly adopted deep learning techniques, particularly convolutional neural networks and Transformer-based architectures\cite{zhao2023towards}. These approaches typically extract specific features from SAR images and compare them directly with class-specific templates learned during training to perform classification. However, such methods inherently rely on the presence of deterministic, explicitly encoded features and often lack the ability to model contextual information or perform reasoning. Feature-based methods may lead to target recognition results that conflict with geographic common sense\cite{geng2021recent}. For example, as shown in Figure 1(a), a large number of ship targets are identified in a port scene, alongside a single aircraft target. This classification is geographically implausible, as aircraft are rarely found in dense dockyard environments where ships dominate the landscape. The presence of the aircraft label in such a context suggests a likely misclassification, indicating a semantic conflict in the recognition output.
Consequently, their performance tends to degrade in scenarios involving ambiguous target structures or strong environmental interference\cite{gao2025recent}.

Moreover, compared to optical imagery, SAR images inherently lack spatial structure and fine-grained texture details\cite{liu2024quality}. As a result, the amount of information presented by targets in SAR images is relatively limited. In many cases, relying solely on a target’s geometric shape or intensity features is insufficient for accurate recognition\cite{zhu2021deep}. Instead, incorporating contextual information, such as the target's location, orientation, and surrounding environment, can significantly enhance recognition performance through reasoning. 
For example, as shown in Figure 1(b), a SAR image of an urban area is provided, in which the top-k candidate target types with confidence scores have been pre-identified by other recognition systems. 
The blue region is located near a major road with convenient access. 
And the relatively large rectangular footprint and high radar reflectivity indicate a complex, multi-level building with hard surfaces. 
So it is more likely to represent a shopping mall. In contrast, the yellow region, being situated within a densely populated city area, is unlikely to be a military base and can therefore be reasonably excluded.
%\begin{figure}[htbp]
%  \centering
%  \includegraphics[width=0.6\textwidth]{1.png} 
%  \caption{An SAR image example of target category inference using contextual information incorporation}
%\end{figure}

\begin{figure}[htbp]
  \centering

  \begin{subfigure}[b]{0.48\textwidth}
    \includegraphics[width=\textwidth]{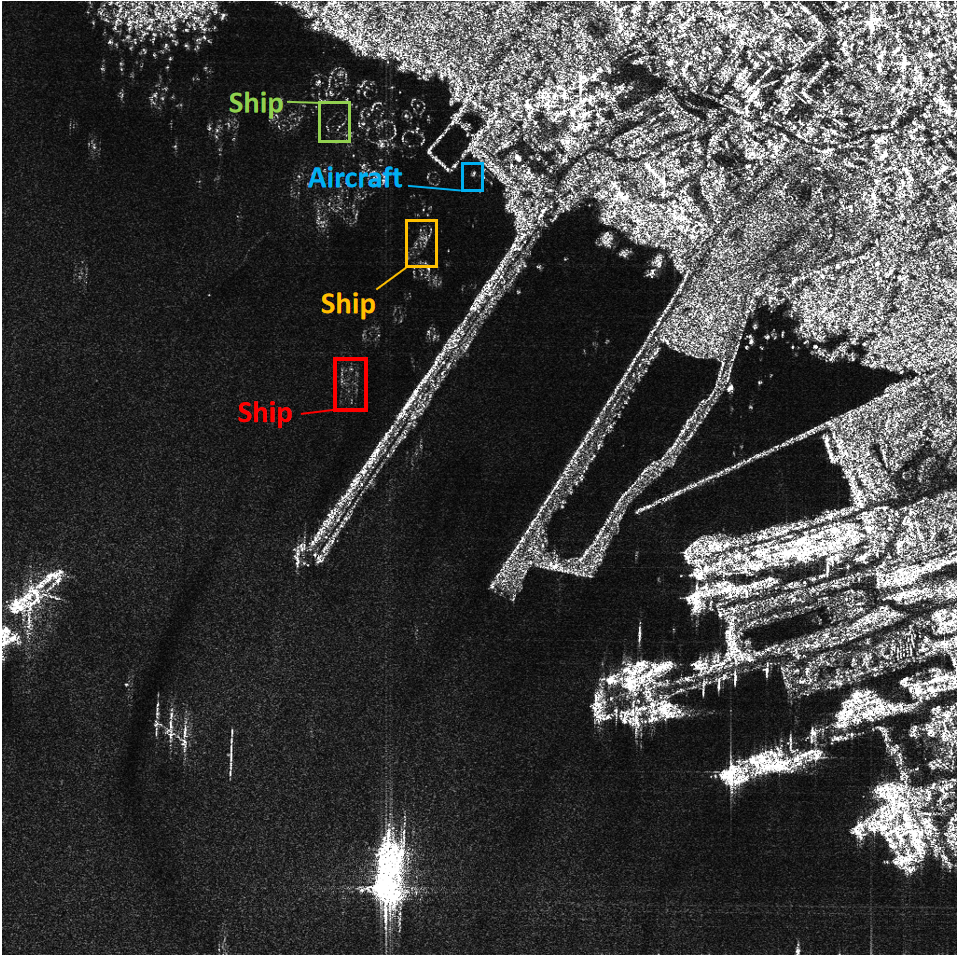}
    \caption{Top-1 prediction from other models causes semantic conflict}
    \label{fig:1a}
  \end{subfigure}
  \hfill
  \begin{subfigure}[b]{0.48\textwidth}
    \includegraphics[width=\textwidth]{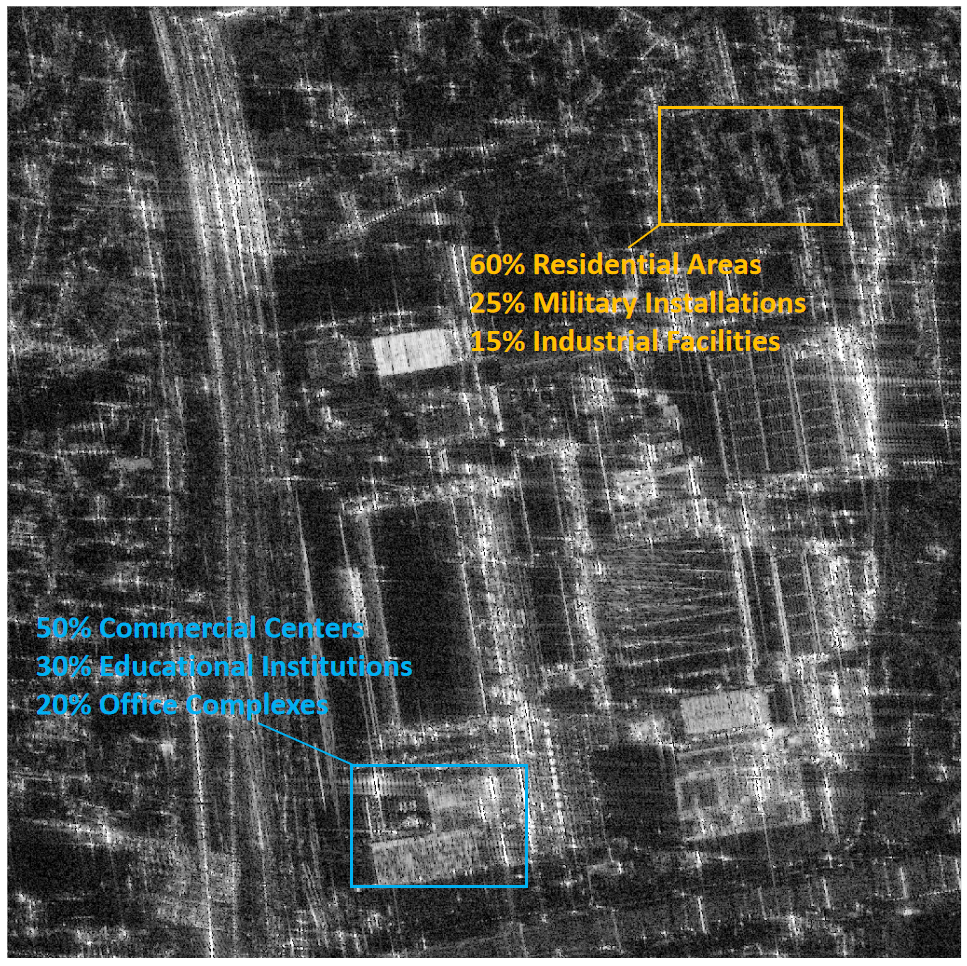}
    \caption{Inferring the target class via reasoning over top-k candidates}
    \label{fig:1b}
  \end{subfigure}

  \caption{Examples of feature-based classification and reasoning-assisted target recognition on SAR imagery}
  \label{fig:comparison}
\end{figure}

In recent years, multimodal large language models have demonstrated strong capabilities in language understanding and multi-step reasoning across various tasks, including natural language processing, code generation, and image interpretation\cite{wu2024multimodal}. 
These models can capture implicit relationships from context and produce logically consistent, interpretable reasoning chains through multi-turn interaction\cite{openai2024gpt4o, google2023gemini}. We argue that such reasoning capabilities have strong potential to be leveraged in SAR image recognition tasks, where they can assist human analysts or automated systems in making informed decisions under ambiguous or incomplete perceptual conditions.

Building on these insights, this work reformulates SAR image recognition as a multi-class reasoning task, where multimodal large language models are employed to infer the most likely target category from a set of candidates and to generate an interpretable chain of thought that records the reasoning process. 
Based on the FAIR-CSAR dataset, we construct a novel SAR reasoning dataset, where each image is paired with  a corresponding reasoning trace generated by a GPT model, along with the final recognition result. 
Preliminary analysis shows that, given candidate options, the GPT model can produce reasonable classification decisions and logically coherent explanations, demonstrating a degree of generalization capability in SAR recognition tasks.

The main contributions of this work are as follows:
\begin{itemize}
  \item SAR image target recognition is reformulated as a multi-class reasoning task guided by a multimodal large language model, with a Chain-of-Thought mechanism introduced to explicitly represent the inference process;
  \item A new dataset for SAR image recognition is constructed, which includes raw SAR images, candidate categories, model-generated reasoning chains, and final recognition results;
  \item The feasibility of applying multimodal large language models to SAR recognition tasks is analyzed, and their potential to assist in decision-making under low-feature or ambiguous conditions is demonstrated.
\end{itemize}

\section{Related Work}
\subsection{SAR Image Target Recognition}
Synthetic Aperture Radar (SAR) imagery has demonstrated significant utility in applications such as military reconnaissance and maritime surveillance, owing to its all-weather, all-day imaging capability. However, due to the intrinsic imaging mechanism of SAR, the resulting images often exhibit high levels of speckle noise, weak texture features, and blurred object boundaries, which hinder the direct application of conventional optical image recognition techniques\cite{zhu2021deep}.

Earlier approaches to SAR target recognition predominantly relied on handcrafted features, such as Histogram of Oriented Gradients (HOG) and Local Binary Patterns (LBP)\cite{gao2020sar}, coupled with classifiers like Support Vector Machines (SVMs). However, these methods suffered from limited feature representation capacity, resulting in suboptimal performance.

In recent years, deep learning has become the dominant paradigm for SAR image recognition. Convolutional Neural Networks (CNNs)\cite{wang2023survey} have demonstrated strong local feature modeling capabilities and have achieved significant progress in target detection and classification tasks. With the rise of Transformer architectures, some studies have introduced lightweight Vision Transformers (ViTs) into SAR image classification pipelines, yielding improved accuracy and generalization across multiple target categories\cite{zhao2023lightViT}. Nevertheless, both CNN and Transformer-based methods rely heavily on explicitly encoded visual patterns in the image. This dependence poses challenges when targets exhibit weak features or structural ambiguity\cite{wu2024faircsar}. Moreover, most existing methods overlook contextual scene information, which is particularly critical in SAR imagery where the visual content of individual targets is often insufficient for accurate recognition.

\subsection{Multimodal Large Language Models and Visual Reasoning Tasks}
In recent years, Multimodal Large Language Models (MLLMs) have achieved breakthrough advancements in tasks such as image understanding, visual question answering, and image captioning\cite{wu2023survey}. By integrating visual and linguistic modalities, these models possess the ability to interpret image semantics from contextual cues and to reason using world knowledge\cite{alayrac2022flamingo}. Representative models such as GPT-4V, Gemini, and Claude have demonstrated outstanding performance in a range of vision-language tasks, particularly in visual reasoning and multimodal dialogue scenarios\cite{wu2023survey}.

In the field of remote sensing, there have been preliminary attempts to apply MLLMs to high-resolution optical remote sensing image understanding, yielding promising results\cite{li2024remotesensingMLLM}. However, to date, there remains a lack of dedicated visual-language reasoning datasets for SAR imagery. Furthermore, there is no comprehensive evaluation of the feasibility of using MLLMs for SAR target recognition tasks, which presents both a gap and an opportunity for future research.

\subsection{Chain-of-Thought Reasoning}
Chain-of-Thought (CoT) is a reasoning paradigm for language models that enhances performance and interpretability on complex tasks by explicitly generating intermediate reasoning steps. This approach was first systematically proposed by Wei et al.\cite{wei2022chain}, and has since been widely adopted in mathematical reasoning, question answering, and multi-hop inference tasks. By decomposing the reasoning process into a sequence of linguistically expressed logical steps, CoT not only improves the reasoning accuracy of language models but also provides human-interpretable insights into model behavior.

In recent years, several benchmark datasets such as GSM8K and SVAMP have incorporated CoT-style annotations to support the training and evaluation of models with step-by-step reasoning capabilities\cite{cobbe2021training,patel2021nlp}. However, in the field of SAR image target recognition, no datasets or task formulations currently leverage the CoT mechanism. Therefore, constructing a SAR image reasoning dataset that includes input images, candidate labels, and CoT reasoning chains is of significant importance for enabling and evaluating large language models in this domain.

\section{Dataset Construction}
\subsection{Data Source}
The SAR imagery used in this study is sourced from the FAIR-CSAR dataset\cite{wu2024faircsar}. The FAIR-CSAR dataset is currently the largest and most detailed fine-grained SAR image dataset available, offering the richest image information and most precise annotations. It covers airports, oil refineries, ports, and riverine regions across 32 locations worldwide, with a total data volume of approximately 250 GB and over 340,000 annotated object instances. The dataset encompasses five major target categories present in SAR images: Aircraft, Ship, Oil tank, Bridge, and Tower crane. Among them, the "Aircraft" category includes 12 sub-classes such as Airbus A220 and Boeing 737, while the "Ship" category includes 7 sub-classes such as Bulk Carrier and Oil Tanker\cite{wu2024faircsar}.

Due to the large scale of the FAIR-CSAR dataset, the current work is conducted on a randomly sampled subset of the data. Additionally, the original dataset includes two ambiguous categories — Other Aircraft and Other Ships, which lack specific semantic definitions. Since it is not feasible to assess the correctness of language model predictions for these categories, we excluded all such instances from our dataset.

\subsection{Data Processing Strategy}
The original FAIR-CSAR dataset includes raw SAR images, text-based target bounding box annotations, and target type labels. We adopted the following strategy to process this data for our task.

First, using the provided bounding box annotations, we overlay a prominent red rectangle at the corresponding position on each SAR image to highlight the target for recognition by open-source multimodal large language models. Since a single SAR image may contain multiple targets (ranging from one to up to approximately 30 targets per image), we generate individual cropped images for each target, with only one bounding box displayed at a time. In other words, each image passed into the large model contains exactly one target for identification.

Next, considering that a large model without access to candidate labels might produce categories that cannot be directly compared with the ground truth, we provide a set of candidate classes for each target, including its true category. This encourages the model to perform inference and choose the most likely category from a constrained set. The candidate list is carefully curated to include visually or semantically similar options. For example, if the true label of a target is the subclass Boeing 737 under the Aircraft category, the distractor classes are selected from other subclasses within the Aircraft category.

Finally, the model is given the SAR image with a single visible target, along with a candidate category list that includes the ground truth. The multimodal large language model performs reasoning based on various forms of knowledge and returns the predicted target category along with its corresponding Chain of Thought explaining the inference process.

\subsection{GPT Interaction Process}
For multimodal language model interaction during data processing, we employed the Azure OpenAI API with the ChatGPT-4o version. In early-stage experiments, some other models such as Gemini are also tested, but ultimately found that ChatGPT performed better on the SAR image recognition task. Therefore, all subsequent reasoning processes in the dataset were generated using the GPT model.

In the prompt design, we introduced confusing options among the candidate categories solely to encourage the model to reason toward the correct category. To prevent the model from relying on elimination strategies to deduce the correct answer by ruling out incorrect candidates, we explicitly instructed GPT not to reason about why the target is not a certain category, but rather to focus only on why it is the predicted correct one.

After the reasoning process is generated by GPT for a given target, the predicted result is evaluated for correctness. Incorrect predictions can arise due to several factors, including targets being too small or partially occluded in the image. To obtain a reliable reasoning chain for each target, any misclassified instance is reprocessed and sent back to GPT for re-inference. For these re-evaluations, we slightly increase the temperature parameter of the GPT model (temperature = 1.2) and modify the candidate categories to differ from those used in the initial attempt. Each such target undergoes up to three rounds of re-inference, and the correct result is retained along with its corresponding reasoning chain.

In rare cases where all three reasoning attempts fail to produce a correct classification, we consider the recognition task for that target beyond GPT’s current capability. These instances are statistically analyzed and reported in Section 4.

All validated reasoning chains are organized by SAR image. For each image, the reasoning processes of all its targets are consolidated into a single data unit, representing the Chain-of-Thought annotations for that image.

\subsection{Data Format}
The dataset consists of two components: SAR image data and corresponding textual annotations. The SAR image data is derived from a carefully curated subset of the FAIR-CSAR dataset. For each SAR image, the textual annotations include: bounding box coordinates of each identifiable target, ground-truth target categories, and GPT-generated chain-of-thought reasoning that outlines the identification process for each target. The target bounding box annotations are derived from the FAIR-CSAR dataset, utilizing the standard polygon bounding box representation format. Each annotation consists of four vertex coordinates arranged as (x1,y1, x2,y2, x3,y3, x4,y4), where the four coordinate pairs correspond to the top-left, top-right, bottom-right, and bottom-left corners respectively. These textual descriptions were systematically compiled to provide comprehensive reasoning support for every SAR image in the dataset.
\begin{figure}[htbp]
  \centering
  \includegraphics[width=0.97\textwidth]{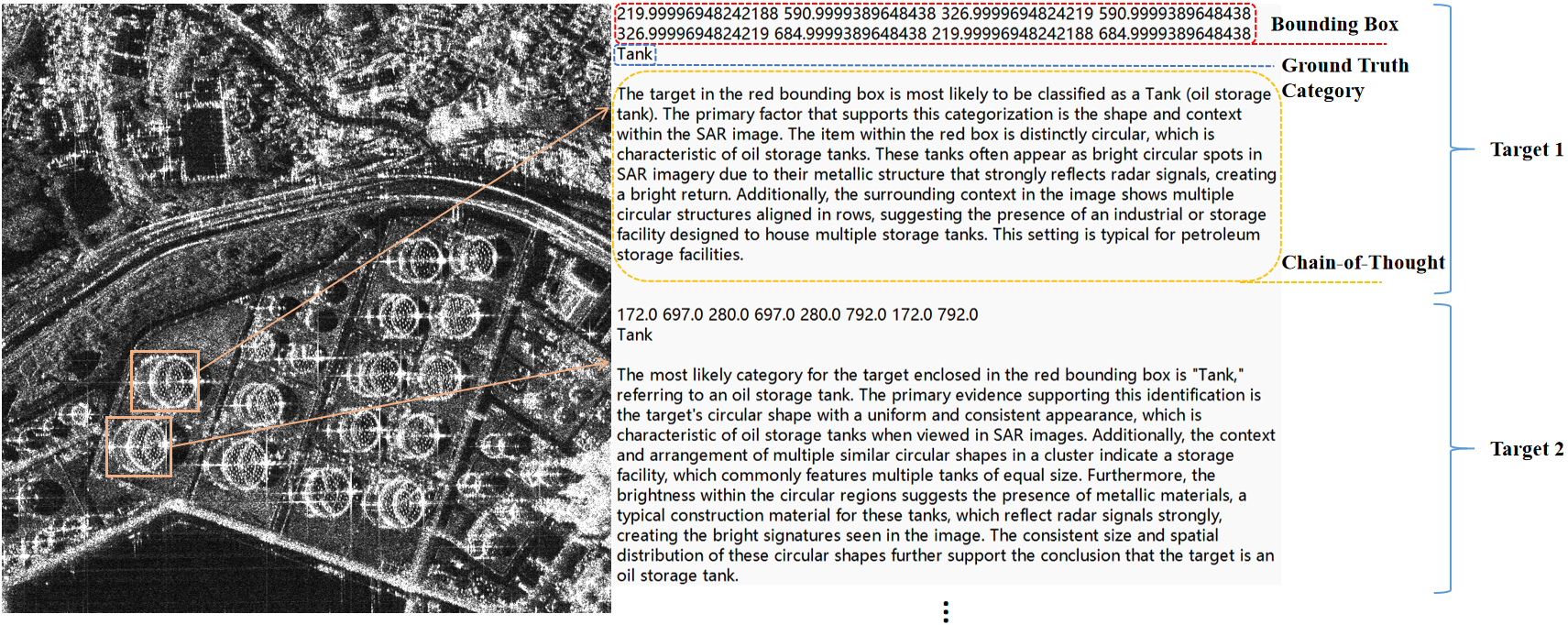} 
  \caption{Data format example in the dataset}
\end{figure}

\section{Dataset Analysis and Statistics}
\subsection{Target Categories and Inference Texts}
A statistical analysis was conducted on the types and quantities of targets present in the SAR images used in this study. The results are shown in Figure 3. Due to the inherent imbalance in category distribution within the original FAIR-CSAR dataset, the sampled data used for this work also exhibits uneven class distribution. Among the categories, General Cargo Ship and Tank account for a relatively high proportion of samples.
\begin{figure}[htbp]
  \centering
  \includegraphics[width=0.78\textwidth]{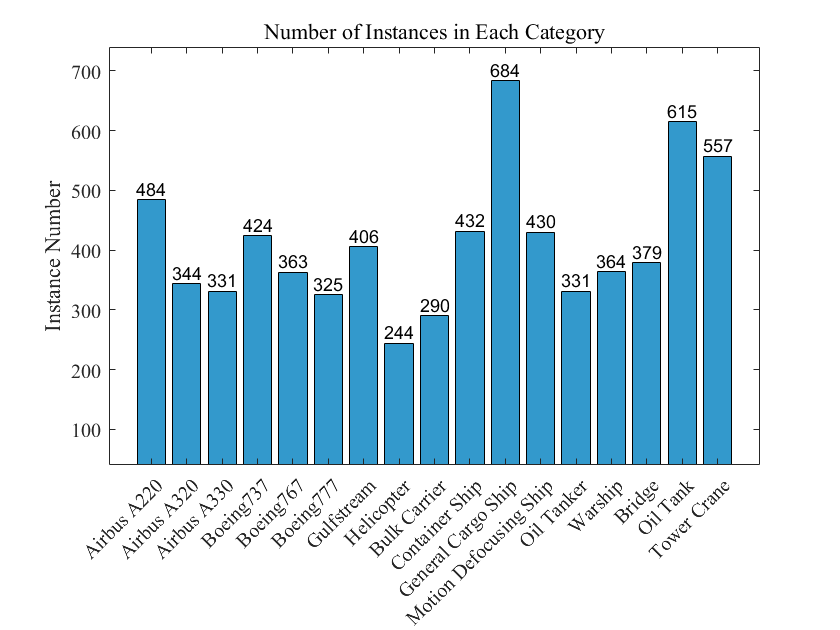} 
  \caption{Number of instances in each category}
\end{figure}

In addition, the length of the reasoning chains generated for each target was analyzed. The reasoning texts produced by GPT typically range from approximately 110 to 200 words. For each inference task submitted to GPT, three candidate categories were provided, consisting of one correct category and two incorrect options.

\subsection{Classification Accuracy Results}
After performing multiple rounds of iterative reasoning, GPT is able to correctly infer the target category for the majority of instances. However, a small subset of targets still yields incorrect predictions even after repeated reasoning attempts. These hard-to-classify instances account for approximately 1.68\% of the total number of targets processed. The misclassified samples are mainly concentrated in five categories: Airbus A220, Gulfstream, General Cargo Ship, Motion Defocusing Ship, and Warship.

The proportion of each category among these misclassified targets is illustrated in Figure 4. Airbus A220 accounts for the largest share, comprising 74.58\% of the total unresolvable cases. This corresponds to 18.18\% of all Airbus A220 samples involved in the reasoning process. For the remaining categories, the misclassification rates relative to their total sample sizes are relatively low, each under 5\%.
\begin{figure}[htbp]
  \centering
  \includegraphics[width=0.82\textwidth]{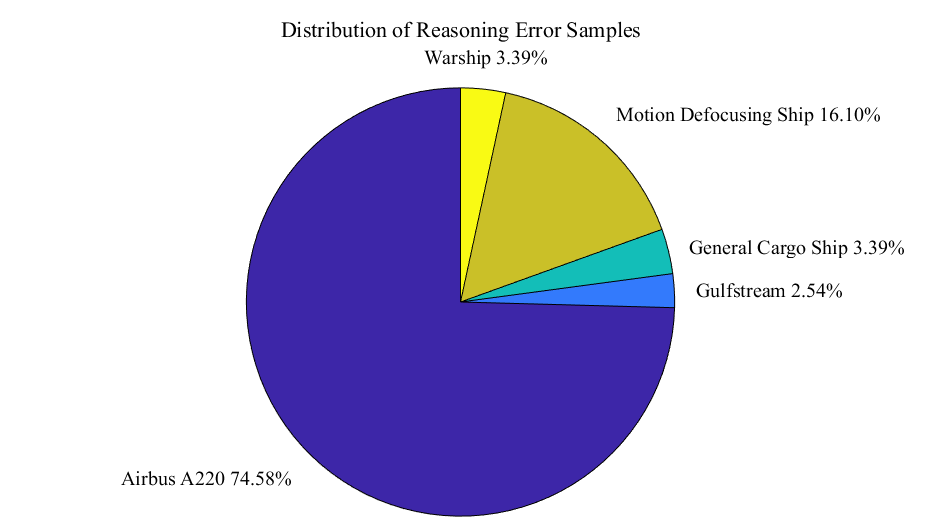} 
  \caption{Distribution of reasoning error samples}
\end{figure}

\subsection{Error Analysis and Model Behavior}
Based on the aforementioned dataset and results, several key issues emerge. One prominent concern is the difficulty in determining whether GPT’s predictions and reasoning processes exhibit signs of overfitting. In certain cases, specific pixel-level features in SAR images, combined with commonsense reasoning, may be applicable to multiple candidate categories. However, GPT often shows a strong preference for one particular category, even when several are plausible. Through repeated reasoning trials, it was observed that GPT may occasionally produce the correct label once, but in aggregate, it tends to generate incorrect predictions more frequently, often following the same erroneous reasoning path.

A closer analysis of specific misclassification patterns reveals that Airbus A220 is the most frequently misclassified category. Interestingly, other similar aircraft types, such as Airbus A320 and Airbus A330, tend to be classified correctly after several reasoning attempts. Airbus A220, however, is frequently misidentified as Airbus A320. This may be attributed to the high visual similarity between these aircraft types, which share many overlapping SAR image features. Given such similarity, GPT appears biased toward categorizing the target as Airbus A320.

Additionally, there are instances where Tower Crane targets were confused with Tank targets during the first reasoning attempt. However, in most cases, this confusion was corrected in subsequent iterations, yielding accurate reasoning outcomes. This suggests that when two categories share some visual similarities but are fundamentally different in function and context, GPT may initially fail to capture all relevant features, but it can refine its reasoning through repeated inference.

\section{Conclusion and Future Work}
In this work, we present a novel Synthetic Aperture Radar (SAR) dataset integrated with open-source multimodal large language model reasoning outputs. The dataset comprises raw SAR imagery encompassing diverse geographical regions, each accompanied by corresponding textual reasoning processes generated by the MLLM for target identification tasks. Experimental validation demonstrates that contemporary open-source MLLMs exhibit measurable SAR image interpretation capabilities, indicating their potential utility as decision-support tools for SAR target classification.

For subsequent research directions, we propose three key developments: First, dataset expansion through increased batch dimensions and more comprehensive candidate option spaces to enhance data diversity and representativeness. Second, implementation of domain expert verification protocols to systematically evaluate the logical validity and technical accuracy of the model's reasoning chains, with particular emphasis on radar-specific knowledge domains such as scattering characteristics. Third, development of fine-tuned multimodal architectures using the validated dataset to optimize model performance for SAR target recognition applications.

%\section*{Acknowledgments}

%Bibliography
\bibliographystyle{unsrt}  
\bibliography{references}

\end{document}